\documentclass[a4paper]{article}
\usepackage[latin2]{inputenc}
\usepackage{icrc2013}
\usepackage{natbib}

\title{Performance of the Cherenkov Telescope Array at energies above 10 TeV}

\shorttitle{CTA at energies above 10 TeV}

\authors{
Anna Barnacka$^{1}$,
Leszek Bogacz$^{2}$,
Mira Grudzi\'nska$^{3}$,
Adam Frankowski$^{4}$,
Mateusz Janiak$^{1}$,
Piotr Lubi\'nski$^{4}$,
Rafa\l{} Moderski$^{1}$
for the CTA Collaboration
}

\afiliations{
$^1$ Nicolaus Copernicus Astronomical Center, Warsaw, Poland \\
$^2$ Jagiellonian University, Krak\'ow, Poland \\
$^3$ Astronomical Observatory of the University of Warsaw, Poland \\
$^4$ Nicolaus Copernicus Astronomical Center, Toru\'n, Poland \\
}

\email{abarnack@camk.edu.pl}

\abstract{The Cherenkov Telescope Array (CTA) is the next generation observatory for very high energy gamma rays. 
	  The capability of the array to detect gamma-rays above 10 TeV is going to be achieved 
	  with a large number of Small Size Telescopes (SSTs) which will cover a large area. 
	  The subarray composed of  SSTs has to compromise the number of telescopes (cost) and the large effective area. 	
	  The separation between the telescopes has to be adjusted to achieve  highest sensitivity 
          with the smallest number of telescopes. 
          On the other hand larger separation can worsen the energy 
          threshold as well as the energy and angular resolutions. 
          In our study we have investigated the optimal spacing between the telescopes of 
          the SST array using an analytical approach and the concept of telescope cell 
          consisting of four telescopes as well as Monte Carlo simulations of the sets of cells.}


\begin{document}
\maketitle

\section{Introduction}
The current imaging atmospheric Cherenkov telescope (IACT) arrays have 
provided a tremendous progress in the field of the very high energy (VHE$>$100~GeV) astronomy.
The latest generation of gamma-ray instruments
like H.E.S.S., MAGIC and
VERITAS consists of two to five telescopes 
equipped with analogue cameras based on photomultiplier tubes. 

The sensitivity of the current instruments reaches the maximum for energies from 100 GeV to 10 TeV.
The best performance in this energy range is obtained by locating 
the telescopes in the inner light pool of the shower ($\sim$120~m from the shower core),
where the distribution of Cherenkov light density is relatively flat. 
The Cherenkov light density at ground level is typically $\sim$10 photons/m$^2$ 
for a shower induced by a 100~GeV $\gamma$-ray event. 
At least 100 photo-electrons (phes) have to be generated to allow  
for the event reconstruction. 
With the optical efficiency of the current instruments 
reaching $\sim$10\%, 
the telescope has to be equipped with at least 
a $\sim$10~m diameter mirror to collect the required number of Cherenkov photons.

The next generation of IACT arrays, the Cherenkov Telescope Array (CTA) \citep{CTA},
is aimed to improve the sensitivity, as compared to the current instruments,
by an order of magnitude in the 100 GeV - 10 TeV energy range and to extend the current sensitivity to lower and higher energies. 

The improvement in the lower energy range is going to be achieved by adding telescopes 
with a large mirror area.
Such telescopes will be able to collect enough Cherenkov photons 
with densities down to $\sim$1 photon/m$^2$.

At higher energies above 10~TeV, the Cherenkov light density in the inner light pool 
is larger than $\sim$1000 photons/m$^2$.
Therefore, even telescopes with small mirror area ($\sim$10~m$^2$) are able 
to collect enough Cherenkov photons to reconstruct  direction and  energy of 
a primary particle. 
But, since source flux rapidly decrease with energy, 
such an array must have a very large effective collection area ($A_{\mathrm{eff}}$).
The CTA, in its design phase, requires the effective collection area to be $>$7~km$^2$
at energies above 10~TeV.
Such a large effective area can be obtained by building dozens of small size telescopes (see e.g. \cite{2008NIMPA.588...48R}).

The Cherenkov light density for distances above $\sim$120~m from the shower core
falls by a factor of $\sim$50 per each 100~m. 
Despite the large density decrease, VHE $\gamma$-ray showers 
could be seen beyond a distance of $\sim$500~m from the shower core. 
The Cherenkov light density of the shower induced by a 100~TeV $\gamma$-ray 
 is still $\sim$100 photons/m$^2$ at a distance of 1000~m from the shower core.

However, the distance between the source position in the camera plane 
and the image center increases with the impact parameter.
An image corresponding to a core distance of 500~m
would be displaced by approximately 4$^\circ$.
Therefore, the detection of distant showers requires 
a camera with a large field of view. 

The subarray composed of  SSTs has to compromise between the number 
of telescopes (cost) and the large effective area.
At fixed cost, the effective area can be increased  by 
setting the telescopes further apart. 
On the other hand, larger separation can worsen the energy 
threshold as well as the energy and angular resolutions.  

The simulations of the whole telescope array are time and resource consuming.  
Therefore, we have used a concept of a telescope cell \citep{1997APh.....6..343A}.
In our study we have investigated the optimal spacing for the array consisting of
four 4m Davies-Cotton Small Size Telescopes.

\section{Cell Concept}
The cell concept has been introduced 
to investigate the performance of the IACTs array at energies above 100~GeV \citep{1997APh.....6..343A}.
A cell is system of four  IACTs located in corners of a rectangle of a given size.
Only events with core distances inside the cell are considered.
The sensitivity of an array comprised of N cells is calculated using results obtained for a single cell,
except effective area for $\gamma$-rays and protons, which are in addition multiplied by number of cells.

For the larger number of telescopes the relative error of the cell approach becomes smaller. 
In the case of cells, an effective area of an array is defined as  $\mathrm{AT_{eff}} \times N_{\mathrm{cells}} \times d^2$, 
where $N_{\mathrm{cells}}$ is a number of grid cells, $d$ is telescope separation and $\mathrm{AT_{eff}}$ 
is an average trigger efficiency in the cell.  
The required number of cells can be derived for given telescope  separation 
and given effective area. 
Lets assume for the moment that $\mathrm{AT_{eff}}$ equals 1.
In such a case, the number of cells, $N_{\mathrm{cells}}$, is defined as:
\begin{equation}
N_{\mathrm{cells}}=\frac{\mathrm{Req\,Effective\,Area}}{d^2}
\label{eq:Ncells}
\end{equation} 

The number of telescopes, $N_{tel}$, in the grid built on square cells is:
\begin{equation}
N_{tel}=(\sqrt{N_{\mathrm{cells}}}+1)^2
\label{eq:Ntel}
\end{equation}

Equations (\ref{eq:Ncells}) and (\ref{eq:Ntel}) can be used to derive the number of 
cells and the number of telescopes for given telescopes separation and required effective area of 7 km$^2$.
These numbers are presented in Table~\ref{Table:EffArea}.

\begin{table}[h]
  \begin{center}
    \begin{tabular}{| c | c | c |}
      \hline \hline
      d [m] & $N_{cells}$ & $N_{tel}$ \\
      \hline \hline
      120 & 529 \\ \hline
      200 & 175 & 202 \\ \hline
      300 & 78 & 96 \\ \hline
      400 & 44 & 58 \\
      \hline
    \end{tabular}
  \end{center}
  \caption{Number of telescopes and cells of a given separation required to reach
    $7\,$km$^2$ effective area.}
  \label{Table:EffArea}
\end{table}

\section{Telescope Parameters \label{sec:telconf}}

The analysis has been performed for a  telescope equipped with a 4~m diameter mirror 
with the focal length of 5.6~m \citep{Moderski}.
The camera consists of 1285 Geiger-mode avalanche photodiodes (G-APDs). 
The physical size of the pixel is 2.32~cm, 
and the field of view~(FoV) of the single pixel is 0.24$^{\circ}$.
The camera has a diameter of 88~cm and FoV of 9$^{\circ}$.

\section{Monte Carlo Simulations and Analysis}

The Monte Carlo (MC) simulations have been performed to examine the performance of the telescope array.  
First, the development of the extensive air showers (EASs) caused by 
$\gamma$-rays and protons is simulated.
The EASs are simulated with {\tt CORSIKA} (COsmic Ray
SImulations for Kascade) \citep{1993AIPC..276..545C} and SIBYLL hadronic interaction model \citep{1994PhRvD..50.5710F}.

The site at an altitude of 2000 m above sea level has been simulated. 
The simulated primary particles entered the atmosphere at zenith angle of 
20$^\circ$. 
The events have been simulated in the energy range from 1~TeV to 100~TeV.
A standard desert atmosphere has been used.

\begin{figure}
  \centering
  \includegraphics[width=0.4\textwidth,angle=-90]{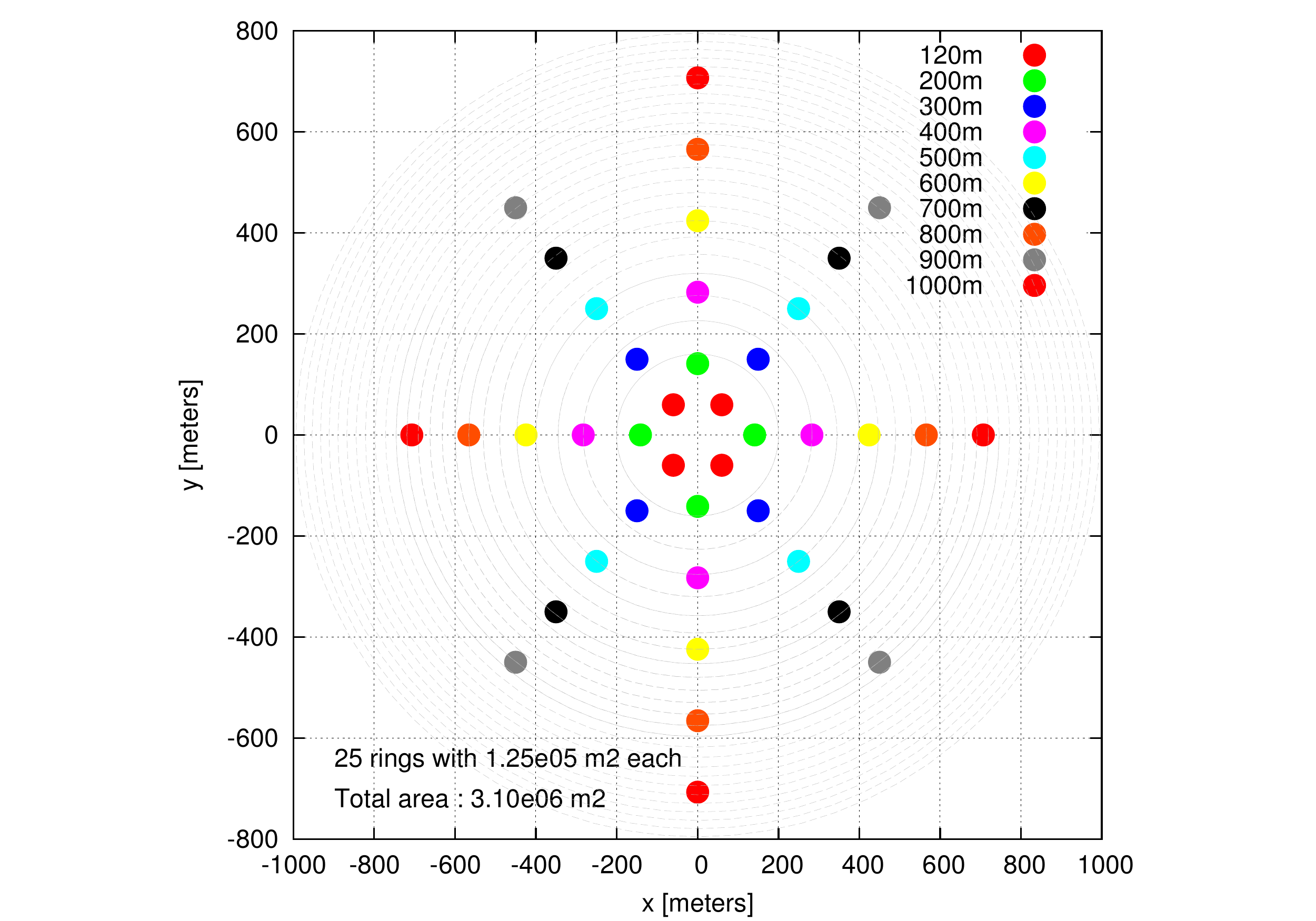}
  \caption{Array layout used in simulations.}
  \label{fig:layout}
 \end{figure}

10 arrays of 4 telescopes have been simulated with a telescope separation
ranging from 120~m up to 1000~m.
The array layout is presented in Figure~\ref{fig:layout}.

We have divided the {\tt CORSIKA} simulations into rings of equal area ($1.25\times10^5$~m$^2$)
and with the same number of events per ring. 
The outer rings have been simulated up to a radius at which 
the trigger efficiency considerably decreased. 
In this simulation the maximum impact parameter was $\sim$1000~m, 
which corresponds to 25 rings. 
  
The ring approach has been chosen to ensure 
sufficient event number at large distances from the array core 
without a loss of computer resources for simulations of events with negligible event detection efficiency.

Next step, the {\tt sim\_telarray} package~
has been used to simulate the telescopes array response to Cherenkov
photons.  
The parameters of the simulated telescope have been given in section \ref{sec:telconf}.

Analysis of {\tt sim\_telarray} output
has been done using {\tt read\_hess} software, which attempts to reconstruct
the events seen by telescopes by applying various image analysis methods,
image cleaning etc \citep{2013APh....43..171B}. 
The results presented here are derived using the so-called ``standard cuts'':
a minimum amplitude of 100 phes has been required, four telescopes had to be triggered, 
at least 5 pixels in the shower image, images have been cleaned using 7/14 value for tail cuts. 
The other cuts on energy and angular resolution have been optimized by simulated annealing method to achieve 
the best sensitivity. 

The sensitivity, an angular resolution and an energy resolution have been calculated 
using definitions given by  \cite{2013APh....43..171B}.
A Boosted Decision Trees method was used to distinguish proton and $\gamma$-ray events. 
Errors were estimated by a bootstrap method.

\section{Results}

 \begin{figure}
  \centering
  \includegraphics[width=0.5\textwidth,angle=0]{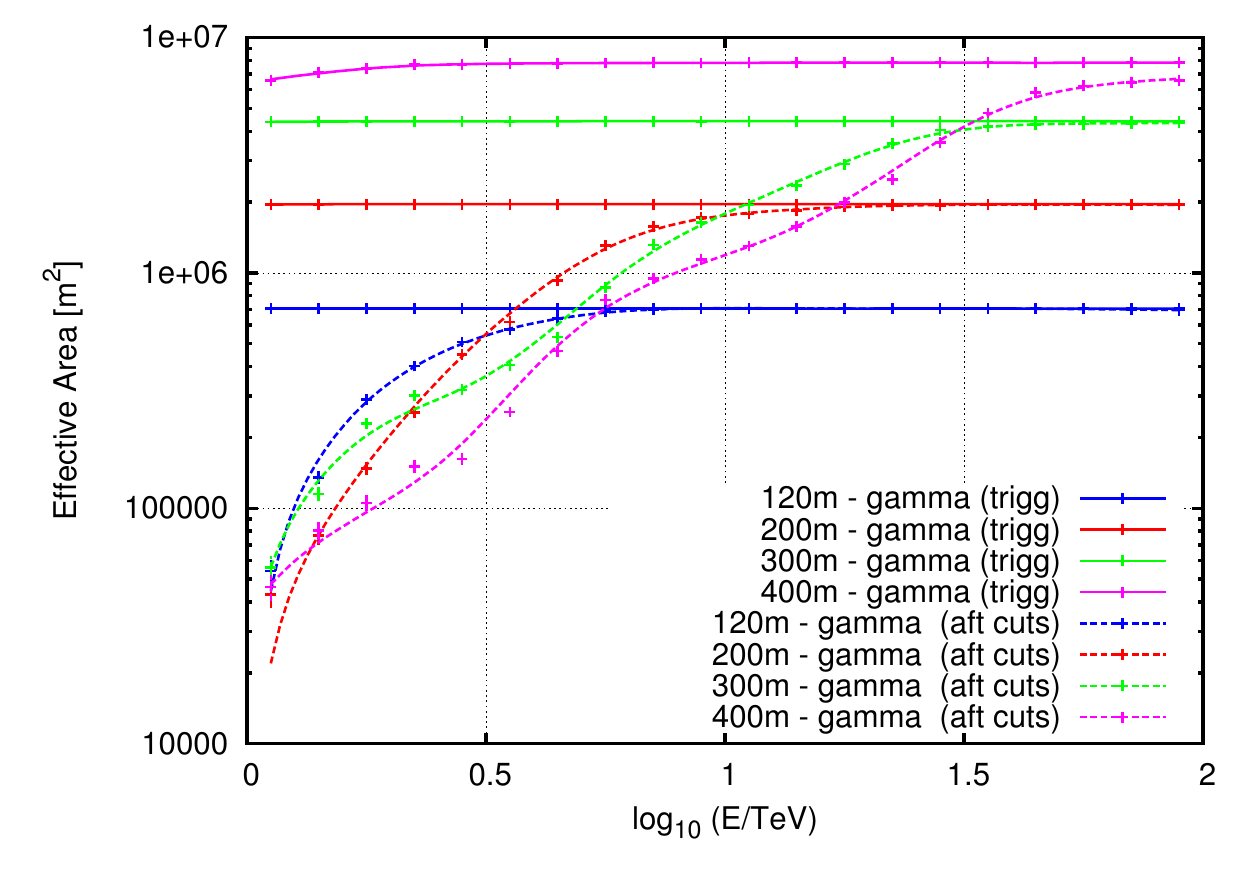}
  \caption{$\gamma$-ray effective area  for array consisting of 49 cells with different separation. 
Effective area has been obtained considering only $\gamma$-ray showers inside the cells. 
The results correspond to effective area after trigger (trigg) and after cuts (aft cuts).}
  \label{fig:eagamma_std}
 \end{figure}

 \begin{figure*}
  \centering
  \includegraphics[width=0.24\textwidth,angle=0]{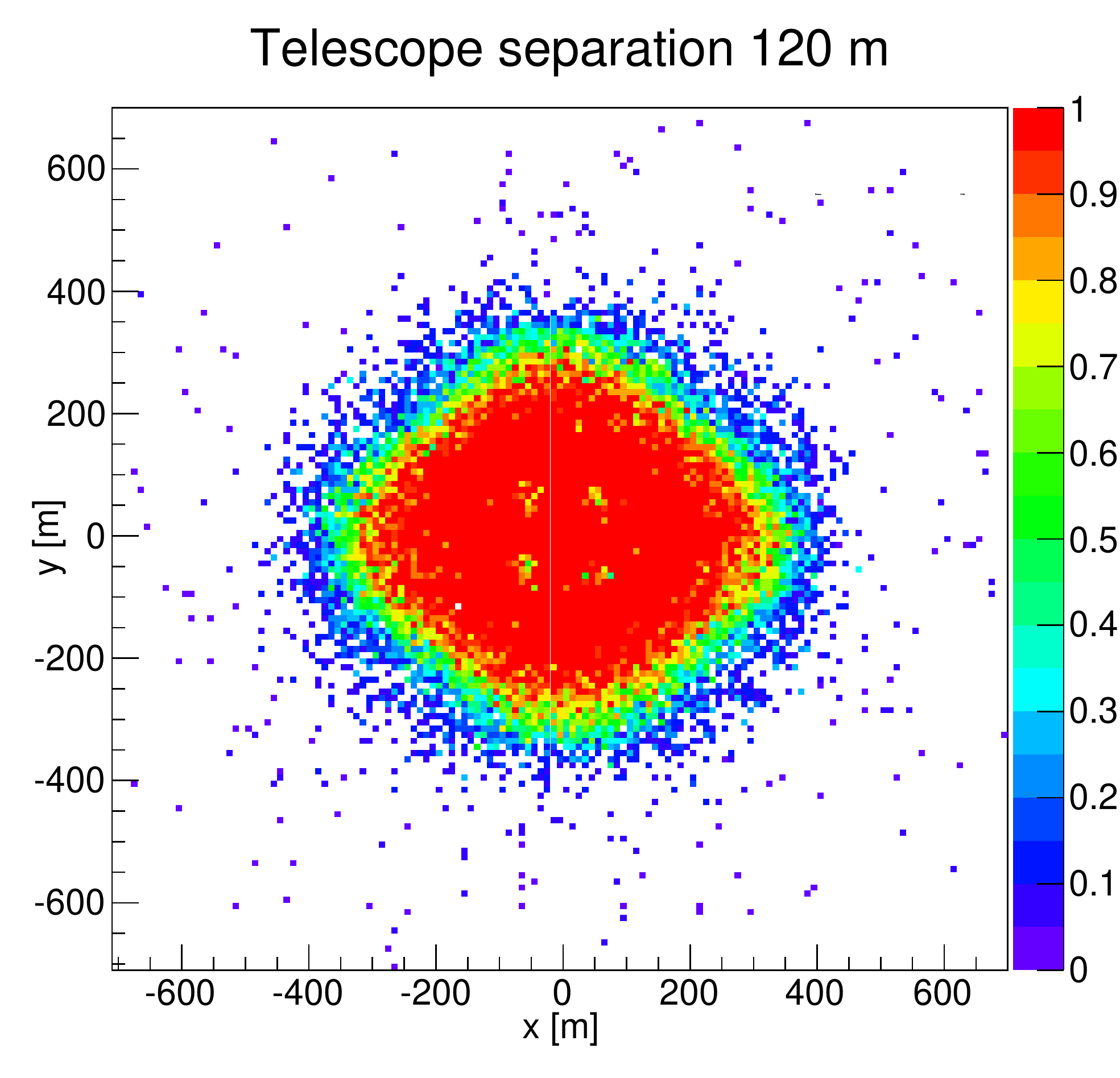}
  \includegraphics[width=0.24\textwidth,angle=0]{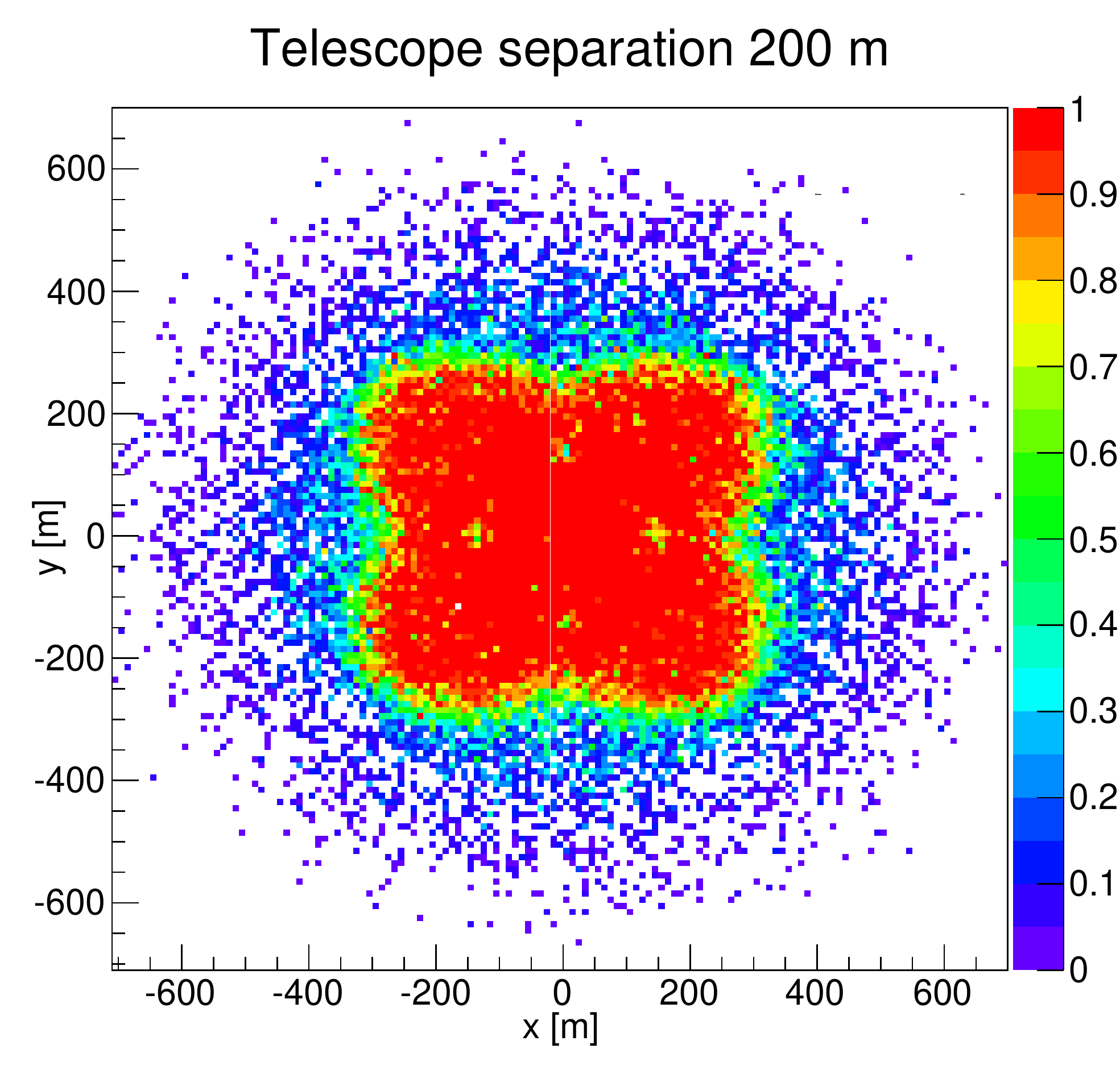}
  \includegraphics[width=0.24\textwidth,angle=0]{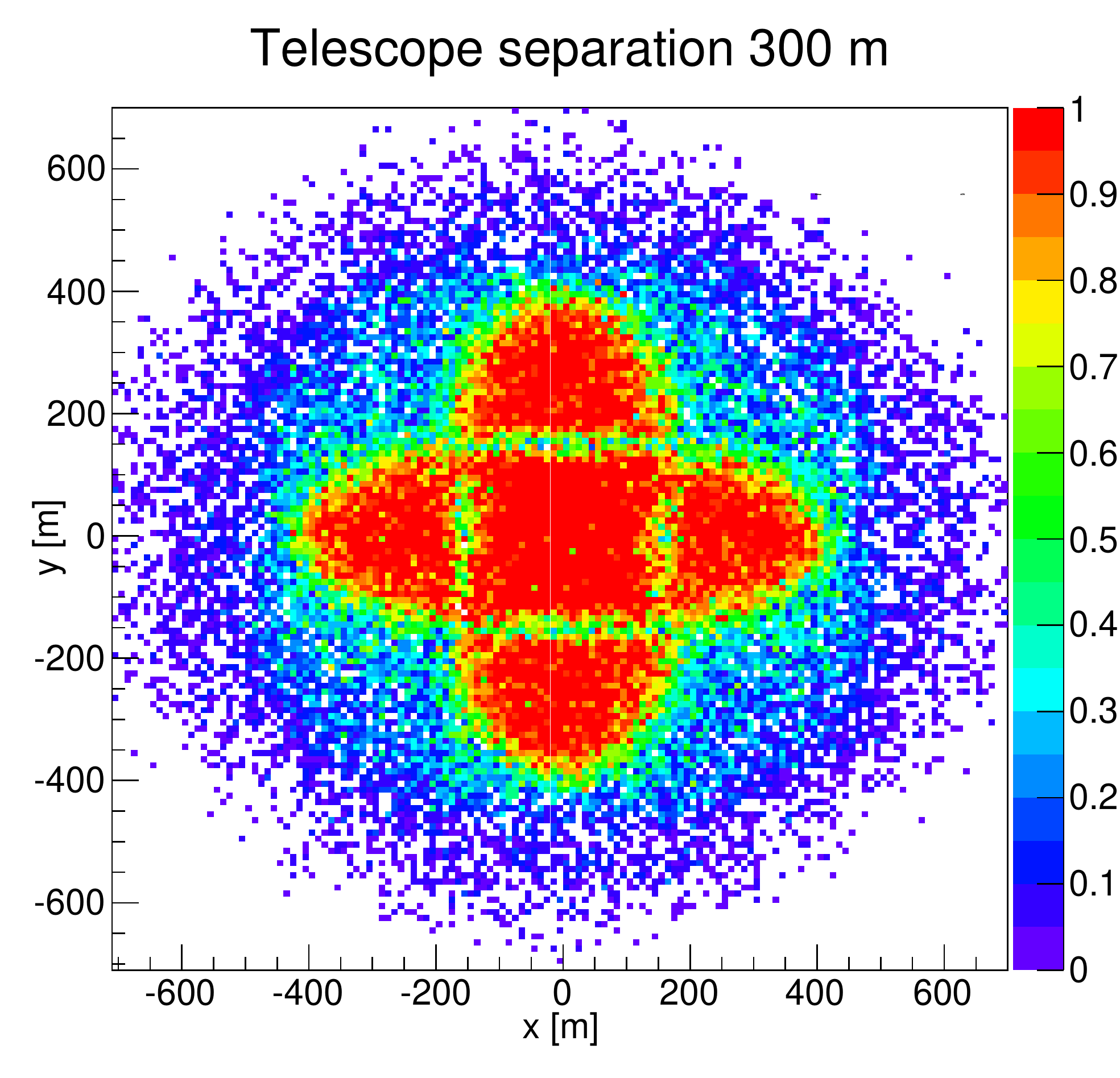}
  \includegraphics[width=0.24\textwidth,angle=0]{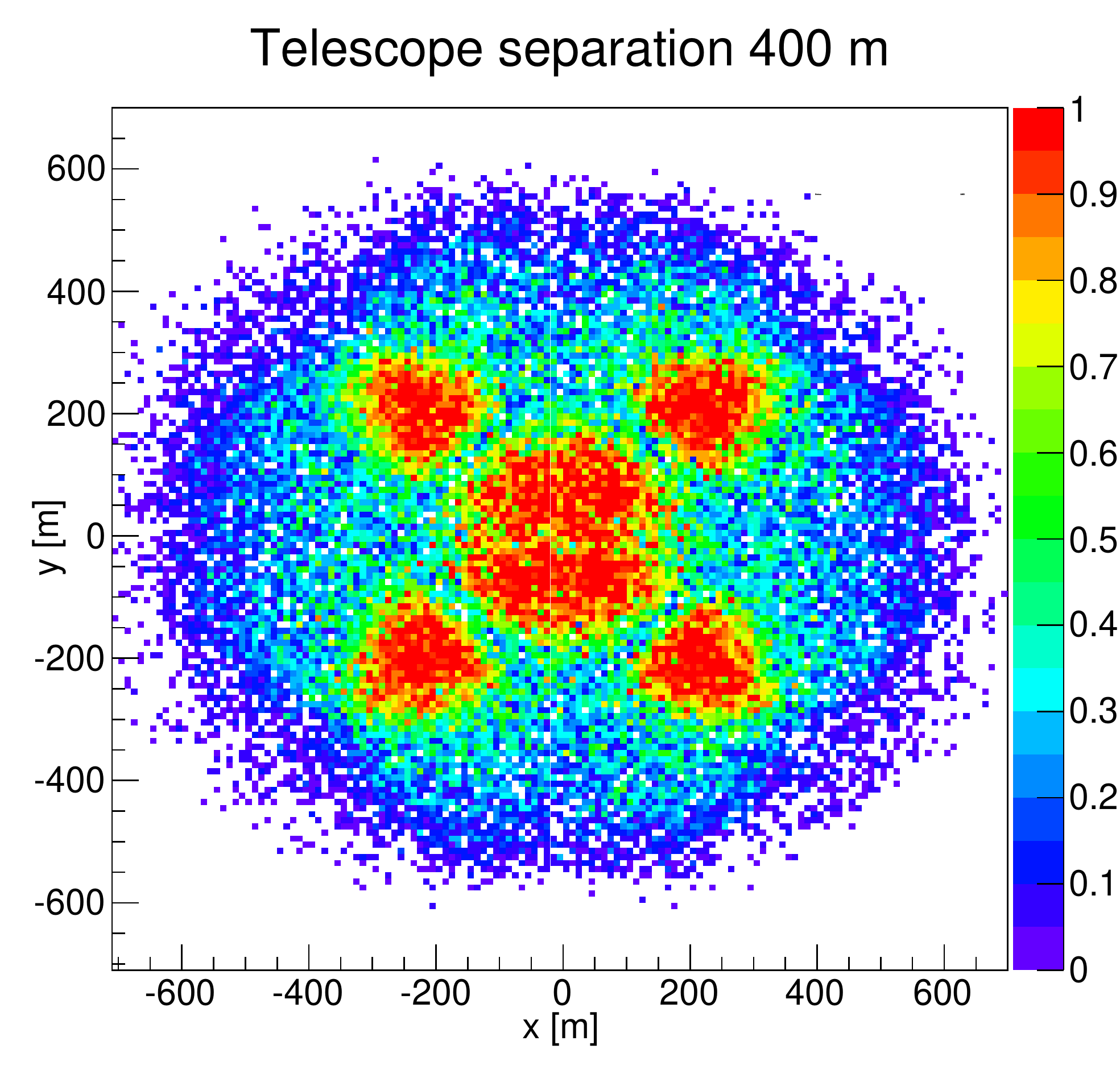}
  \caption{Trigger efficiency of all events (also outside the cell) after cuts for different telescope separations: 
           120~m, 200~m, 300~m and 400~m, respectively. 
           Figures represent $\gamma-ray$ events with energy $\sim$100~TeV. }
  \label{fig:TriggerEff100TeV}
 \end{figure*}

In this paper, the results of the analysis of square four-telescope cells
with side lengths of 120~m, 200~m, 300~m and 400~m are presented.
The results for distances above 400~m have been omitted 
due to the significantly worse sensitivities. 
The final sample of simulated events consists of 11.6 mln of $\gamma$-ray events 
and 33.2 mln of proton events. 

The results obtained for four-telescope cells have then  been used to extrapolate 
the sensitivity of an array of 64 telescopes corresponding to 49 cells. 

Figure~\ref{fig:eagamma_std} shows the effective area of the 49 cells. 
The simulations imply that almost all $\gamma$-ray events inside the cell are triggered. 
A significant fraction of the events is lost after applying cuts.
Figure~\ref{fig:TriggerEff100TeV} presents trigger efficiency after cuts 
as a function of the event location. 
The trigger efficiency is shown for events at energy of $\sim$100~TeV.
It is confirmed that even after cuts the very high energy$\gamma$-ray 
events can be detected up to distances above 500~m from the shower core.

\begin{figure}
  \centering
  \includegraphics[width=0.5\textwidth,angle=0]{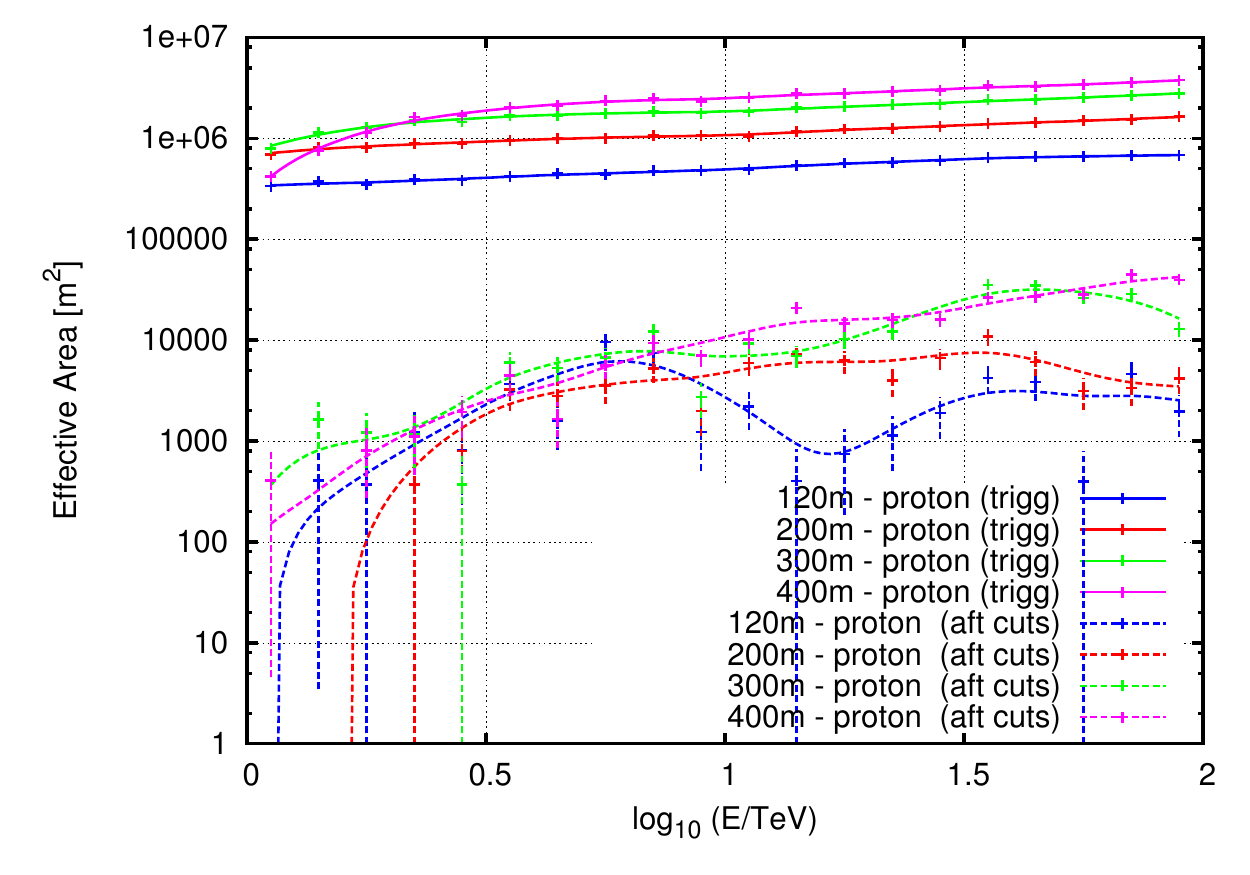}
  \caption{Proton effective area for array consisting of 49 cells with different length sides. 
Effective area has been obtained considering only proton induced showers inside the cells.
The results corresponds to effective area after trigger (trigg) and after all  cuts (aft cuts).  }
  \label{fig:protons_std}
 \end{figure}
The effective area of the array for proton events have also been investigated (see Figure~\ref{fig:protons_std}). 
In the energy range from 1~TeV to 100~TeV almost all protons  inside the cell are detected. 
However, the applied analysis  provides very good proton rejection.
The proton effective area after cuts is reduced by a factor of $\sim$100 for energies above 10~TeV.

 \begin{figure}
  \centering
  \includegraphics[width=0.5\textwidth,angle=0]{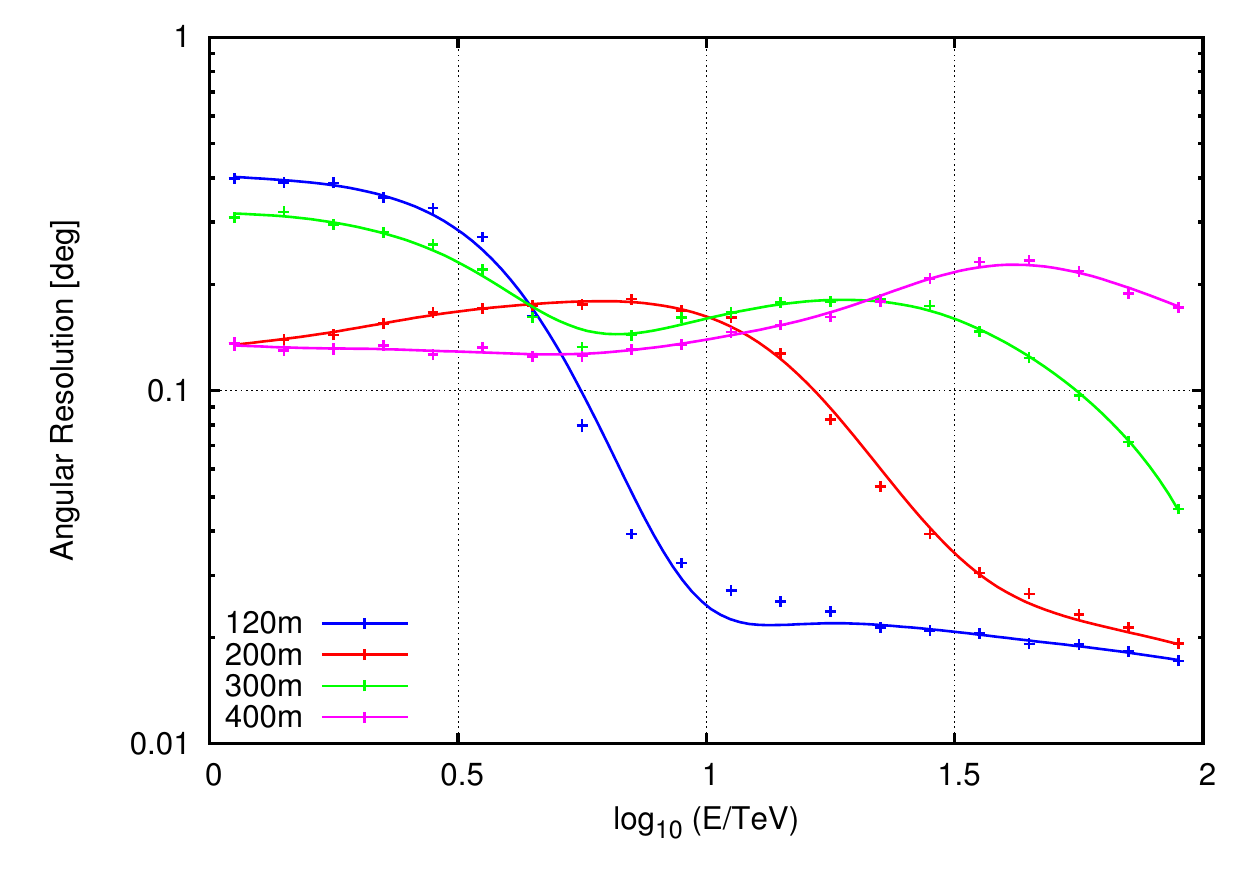}
  \caption{Angular resolution as a function of energy for events inside the cells.
 Cuts were optimized to get best sensitivity.}
  \label{fig:angres_std_all}
 \end{figure}

 \begin{figure}
  \centering
  \includegraphics[width=0.5\textwidth,angle=0]{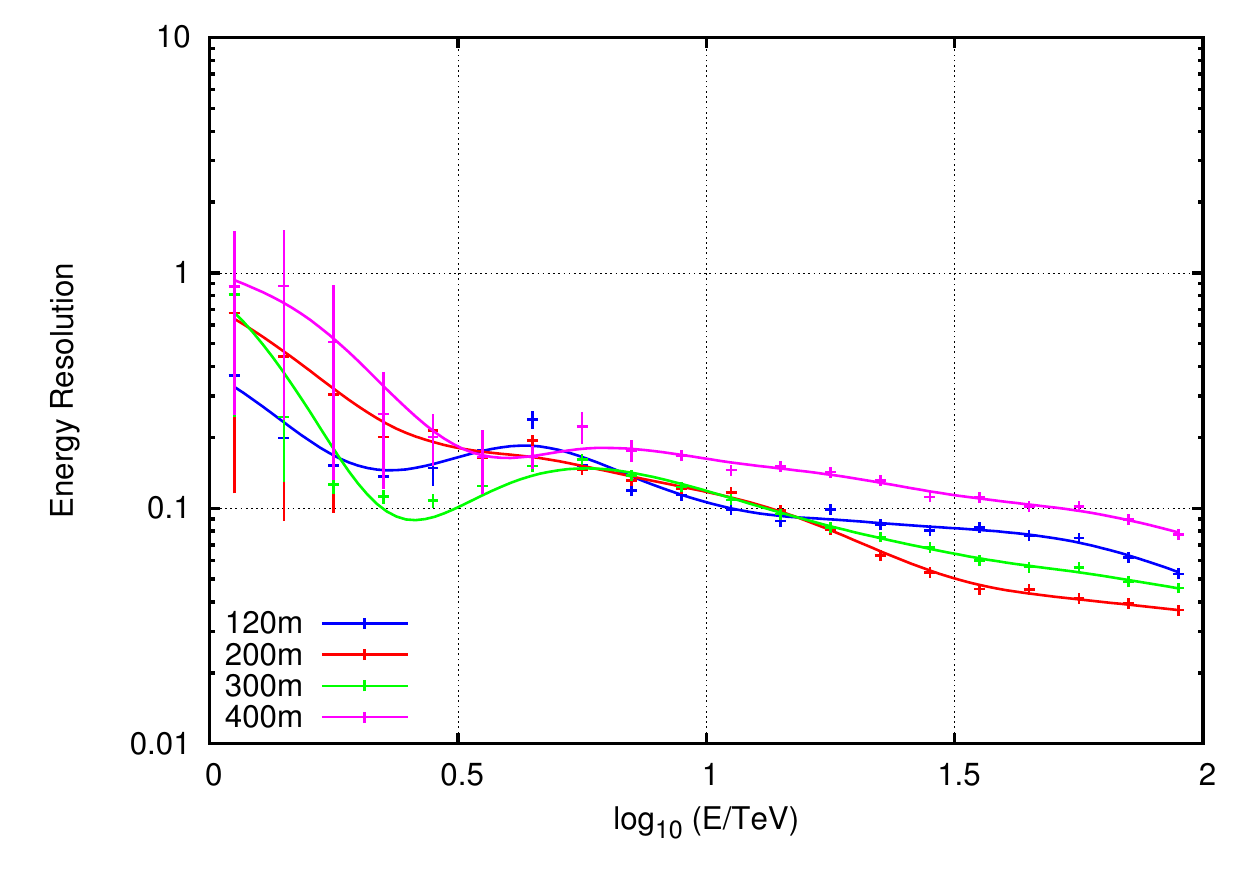}
  \caption{Energy resolution as a function of energy for events inside the cells.}
  \label{fig:eneres_std}
 \end{figure}

\begin{figure}
  \centering
  \includegraphics[width=0.5\textwidth,angle=0]{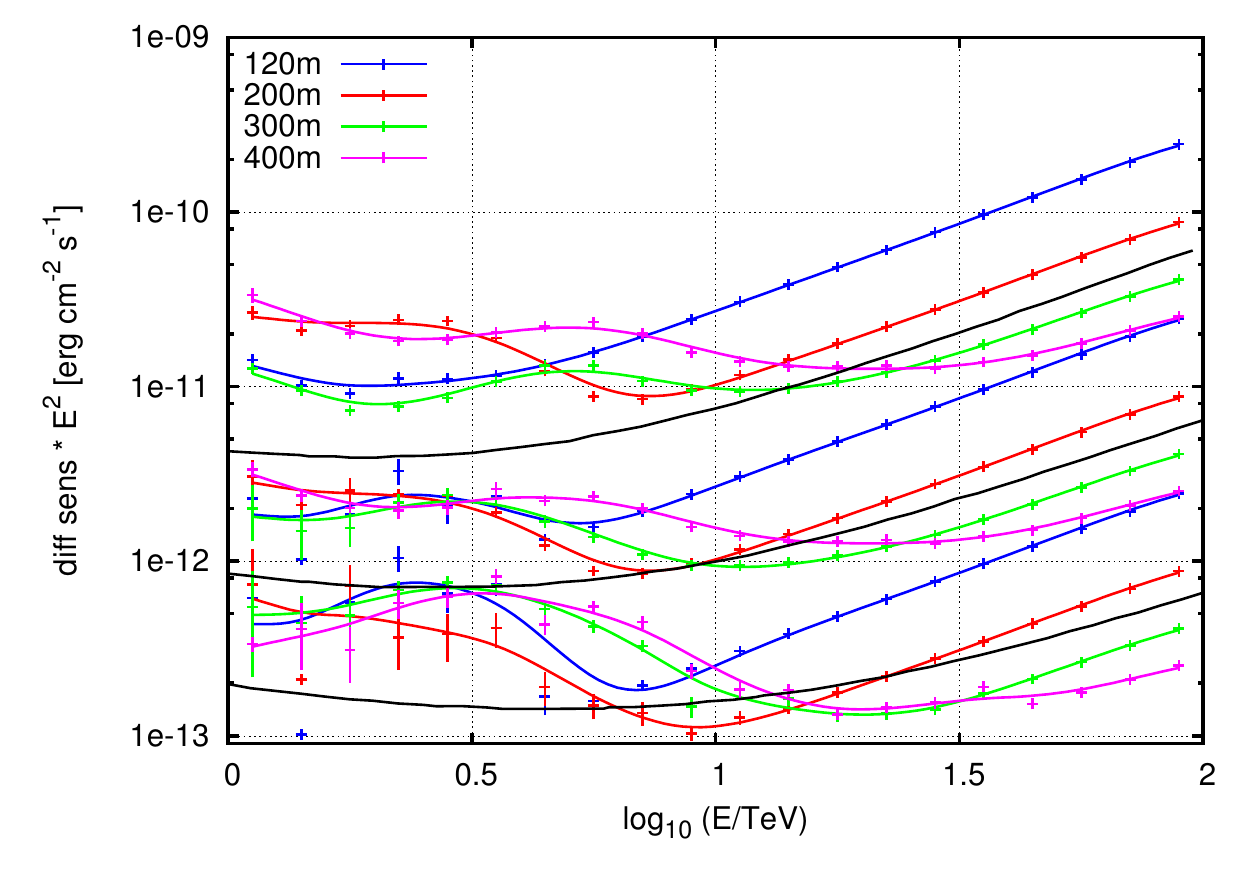}
  \caption{The expected point source differential sensitivity of 
           the 64-telescope array for a different telescope separation. 
	   The solid black lines indicate the required sensitivity of the array.
	   Sensitivity of the array is shown for observation  times of 0.5 hours, 5 hours, and 50 hours, from top to bottom. }
  \label{fig:sens}
 \end{figure}

Angular resolution and energy resolution are shown on Figures~\ref{fig:angres_std_all} and \ref{fig:eneres_std}, respectively.
Finally, Figure~\ref{fig:sens} presents differential point source sensitivity 
for observation times of 0.5 hours, 5 hours, and 50 hours.
The sensitivity of 49 cells array is shown together with the sensitivity required by the CTA.

\section{Conclusions and Discussion}

The currently existing analysis methods have been optimized to work well 
in the energy range from 100~GeV to 10~TeV,
and have been widely used by experiments like H.E.S.S., Veritas or MAGIC. 
New methods optimized for observations above 10~TeV would provide in the future 
a significant improvement in the array performance.

The study presented here shows that the sensitivity of SST array can be substantially
improved in the regime above 10~TeV
through the increased telescope separation. 
The sensitivity for the SSTs is constrained by proton background for energies below 10 TeV.
At higher energies, sensitivity is only constraint by a number of $\gamma$-ray events. 
The improvement in the sensitivity is achieved by enhancing the effective area.
The cell approach shows that the sensitivity required by the CTA in energies above 10~TeV
is met with 64 SST 4~m~DC telescopes, with spacing above 200~m.
  
Figure~\ref{fig:eagamma_std} shows that a significant fraction of the events 
with energies below 10~TeV is lost after all cuts. 
The number of remaining events  at the energies above 10~TeV tends to the number of triggered events.

The obtained angular and energy resolutions of the array depend strongly on the detailed analysis method.
The cuts used here have been optimized to achieve the best sensitivity despite of energy resolution. 

In this study we did not use the timing information.
It has been shown by \cite{2011APh....34..886S} that the angular resolution 
can be significantly improved by using timing gradient.
The background rejection also  can be enhanced 
by introducing telescope timing characteristic \citep{2006APh....26...69D}.

The adopted cell approach constrains analysis only to events inside the cell. 
In reality showers will be detected by a larger number of telescopes. 
The cell concept provides then the ``safe'' approximation - 
only the lower limits on the array performance.

The further improvement of the results is going to be achieved 
by additional proton simulation to increase the statistic. 
Also $\gamma$-rays in the energy range from 100~TeV up to 300~TeV 
will be simulated. 
We also plan  to improve the analysis methods by adding the timing information. 
We will investigate the accuracy of the cell approach using ongoing Prod2 simulation. 
The aim is to compare the results of full SST array with given spacing  to an array built of cells with similar spacing to one used in Prod2.

\vspace*{0.5cm}
\footnotesize{{\bf Acknowledgment:}{We gratefully acknowledge support from
the agencies and organizations listed in this page: http://www.ctaobservatory.org/?q=node/22. The work has been supported by
the National Centre for Research and Development through
project ERA-NET-ASPERA/01/10, the National Science Centre through project DEC-2011/01/M/ST9/01891, and Ministry
of Science and Higher Education project 498/FNiTP/158/2010.
This research was also supported in part by PL-Grid infrastructure.}}

\bibliographystyle{plainnat}
\bibliography{icrc2013-387}

\end{document}